# GW151226: Observation of Gravitational Waves from a 22-Solar-Mass Binary Black Hole Coalescence

B. P. Abbott *et al.*[*]

(LIGO Scientific Collaboration and Virgo Collaboration)

We report the observation of a gravitational-wave signal produced by the coalescence of two stellar-mass black holes. The signal, GW151226, was observed by the twin detectors of the Laser Interferometer Gravitational-Wave Observatory (LIGO) on December 26, 2015 at 03:38:53 UTC. The signal was initially identified within 70 s by an online matched-filter search targeting binary coalescences. Subsequent off-line analyses recovered GW151226 with a network signal-to-noise ratio of 13 and a significance greater than $5\sigma$. The signal persisted in the LIGO frequency band for approximately 1 s, increasing in frequency and amplitude over about 55 cycles from 35 to 450 Hz, and reached a peak gravitational strain of $3.4^{+0.7}_{-0.9} \times 10^{-22}$. The inferred source-frame initial black hole masses are $14.2^{+8.3}_{-3.7} M_\odot$ and $7.5^{+2.3}_{-2.3} M_\odot$, and the final black hole mass is $20.8^{+6.1}_{-1.7} M_\odot$. We find that at least one of the component black holes has spin greater than 0.2. This source is located at a luminosity distance of $440^{+180}_{-190}$ Mpc corresponding to a redshift of $0.09^{+0.03}_{-0.04}$. All uncertainties define a 90% credible interval. This second gravitational-wave observation provides improved constraints on stellar populations and on deviations from general relativity.

## I. INTRODUCTION

A century after Einstein predicted the existence of gravitational waves [1], the Laser Interferometer Gravitational-Wave Observatory (LIGO) [2,3] observed the first gravitational-wave signal GW150914 from a binary black hole merger [4]. In this Letter, we report the observation of a second coincident signal GW151226, also from the coalescence of two black holes. An analysis of GW150914 and GW151226 as a population is described in [5]. LVT151012, the third most significant binary black hole candidate, is also included in this analysis (see Fig. 2 below). No other significant binary black hole candidates in the total mass range $4$–$100 M_\odot$ were found during Advanced LIGO's first observing period, September 12, 2015 to January 19, 2016.

Matched filtering [6–12] was essential to the detection of GW151226 since the signal has a smaller strain amplitude and the detectable signal energy is spread over a longer time interval than GW150914. Detection [13–18] and parameter estimation [19–21] rely on understanding the sources of detector noise [22,23] and on precise waveform models of compact binary coalescence. Waveforms have been developed combining various techniques to model the two-body dynamics and gravitational waves, notably post-Newtonian theory [24–28], the effective-one-body formalism [29–33], and numerical relativity [34–39]. Matched filtering correlates a waveform model with the data over the detectors' sensitive band, which enabled GW151226 to be extracted from the detector noise.

## II. OBSERVATION

On December 26, 2015, the gravitational-wave candidate GW151226 was identified within 70 s by an online matched-filter search [17]. The candidate had an inferred coalescence time of 03:38:53.647 UTC at LIGO Livingston and $1.1^{+0.3}_{-0.3}$ ms later at LIGO Hanford. False alarms more significant than GW151226 would, in principle, be produced by the online search at a rate of approximately 1 per 1000 yr. The candidate signal thus passed the threshold for generating an alert to electromagnetic partners [40]. The source was localized to $\sim$1400 deg$^2$ on the sky (90% credible level) within 3 min of the initial observation [41]. Coarse sky localization is due to the limited information afforded by only two sensitive detectors in observing mode. The initial identification of this signal was confirmed by performing two independent off-line matched-filter searches [14,17,18] that used the waveform models in Refs. [42,43]. Both searches identified GW151226 as a highly significant event. Because of the signal's smaller strain amplitude and time-frequency morphology, the generic transient searches that initially identified GW150914 [44] did not detect GW151226.

Based on current waveform modeling, we find that GW151226 passed through LIGO's sensitive band in 1 s, increasing in frequency over approximately 55 cycles from 35 Hz to a peak amplitude at 450 Hz. The signal-to-noise ratio (SNR) accumulates equally in the early inspiral ($\sim$45 cycles from 35 to 100 Hz) and late inspiral to merger ($\sim$10 cycles from 100 to 450 Hz). This is different from the more massive GW150914 binary for which only the last 10 cycles, comprising inspiral and merger, dominated the SNR. As a consequence, the parameters characterizing GW151226 have different precision than those of GW150914. The chirp mass [26,45], which controls the binary's evolution during the early inspiral, is determined very precisely. The individual masses, which rely on information from the late inspiral and merger, are measured far less precisely.

[*]Full author list given at end of the article.

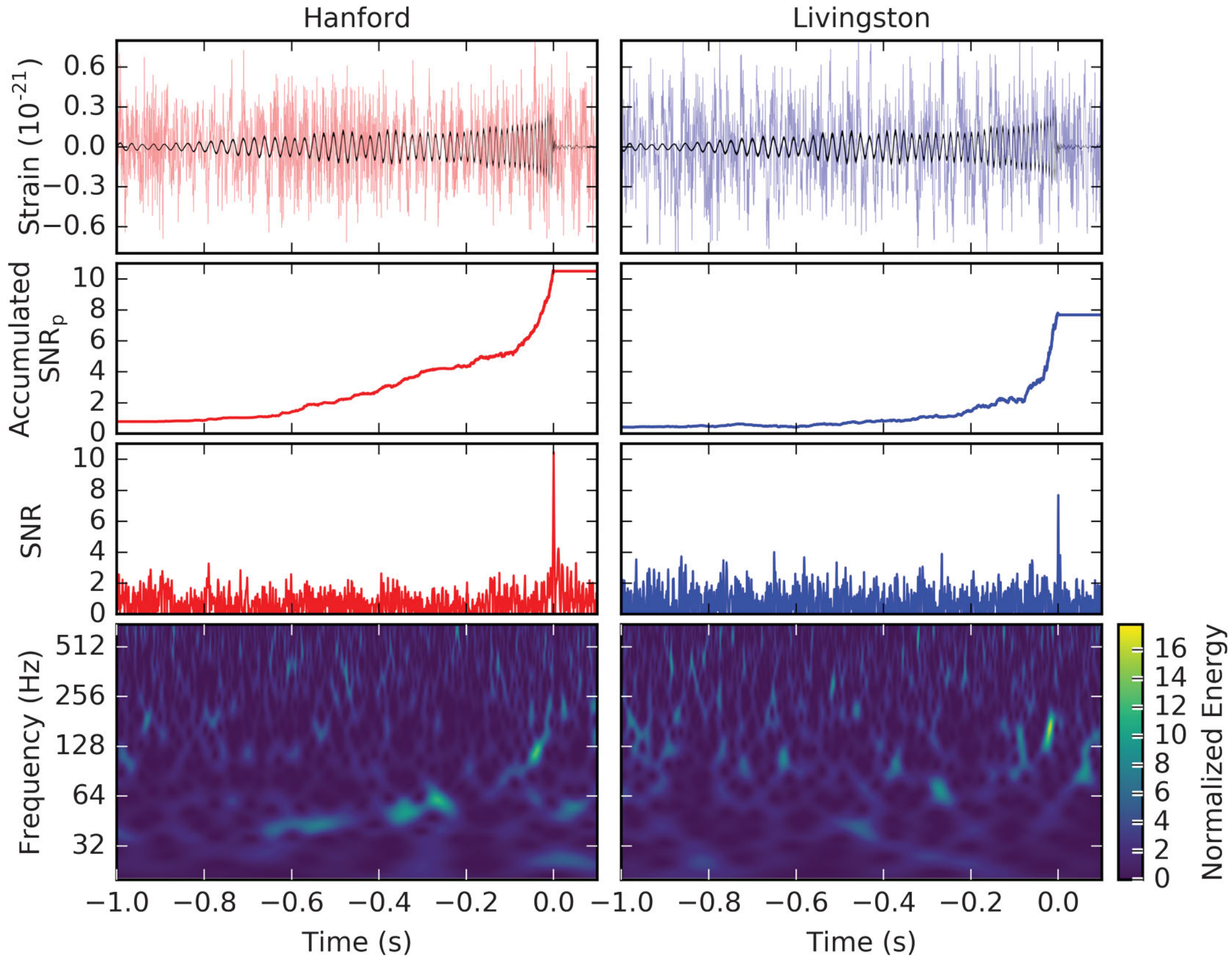

FIG. 1. GW151226 observed by the LIGO Hanford (left column) and Livingston (right column) detectors, where times are relative to December 26, 2015 at 03:38:53.648 UTC. *First row:* Strain data from the two detectors, where the data are filtered with a 30–600-Hz bandpass filter to suppress large fluctuations outside this range and band-reject filters to remove strong instrumental spectral lines [46]. Also shown (black) is the best-match template from a nonprecessing spin waveform model reconstructed using a Bayesian analysis [21] with the same filtering applied. As a result, modulations in the waveform are present due to this conditioning and not due to precession effects. The thickness of the line indicates the 90% credible region. See Fig. 5 for a reconstruction of the best-match template with no filtering applied. *Second row:* The accumulated peak signal-to-noise ratio ($\mathrm{SNR_p}$) as a function of time when integrating from the start of the best-match template, corresponding to a gravitational-wave frequency of 30 Hz, up to its merger time. The total accumulated $\mathrm{SNR_p}$ corresponds to the peak in the next row. *Third row:* Signal-to-noise ratio (SNR) time series produced by time shifting the best-match template waveform and computing the integrated SNR at each point in time. The peak of the SNR time series gives the merger time of the best-match template for which the highest overlap with the data is achieved. The single-detector SNRs in LIGO Hanford and Livingston are 10.5 and 7.9, respectively, primarily because of the detectors' differing sensitivities. *Fourth row:* Time-frequency representation [47] of the strain data around the time of GW151226. In contrast to GW150914 [4], the signal is not easily visible.

Figure 1 illustrates that the amplitude of the signal is less than the level of the detector noise, where the maximum strain of the signal is $3.4^{+0.7}_{-0.9} \times 10^{-22}$ and $3.4^{+0.8}_{-0.9} \times 10^{-22}$ in LIGO Hanford and Livingston, respectively. The time-frequency representation of the detector data shows that the signal is not easily visible. The signal is more apparent in LIGO Hanford where the SNR is larger. The SNR difference is predominantly due to the different sensitivities of the detectors at the time. Only with the accumulated SNR from matched filtering does the signal become apparent in both detectors.

## III. DETECTORS

The LIGO detectors measure gravitational-wave strain using two modified Michelson interferometers located in Hanford, WA and Livingston, LA [2,3,46]. The two orthogonal arms of each interferometer are 4 km in length, each with an optical cavity formed by two mirrors acting as test masses. A passing gravitational wave alters the differential arm length so that the measured difference is $\Delta L(t) = \delta L_x - \delta L_y = h(t)L$, where $L = L_x = L_y$ and $h$ is the gravitational-wave strain amplitude projected onto the detector. Calibration of the interferometers is performed by inducing test mass motion using photon pressure from a modulated calibration laser. Employing methods as described in [48], the calibration uncertainty ($1\sigma$) in both detectors at the time of the signal is better than 8% in amplitude and 5 deg in phase.

At the time of GW151226, both LIGO detectors were operating with a sensitivity typical of that exhibited throughout the observing period [46]. Investigations similar to the detection validation procedures for GW150914 found no evidence that instrumental or environmental disturbances contributed to GW151226 [4,23]. Tests quantifying the detectors' susceptibility to external environmental disturbances, such as electromagnetic fields [49], indicated that any disturbance strong enough to account for the signal would be clearly detected by the array of environmental sensors. All environmental fluctuations recorded during GW151226 were too small to account for more than 6% of its peak strain amplitude. Furthermore, none of the environmental sensors recorded any disturbances that evolved in time and frequency like GW151226.

## IV. SEARCHES

Two matched-filter searches [18] used coincident observations between the two LIGO detectors from September 12, 2015 to January 19, 2016 to estimate the significance of GW151226 [5]. One of these searches was the off-line version of the online search discussed previously [17]. The off-line searches benefit from improved calibration and refined data quality information not available to online searches [5,23].

Each search identifies coincident events that are found in both LIGO detectors with the same template and within 15 ms [18]. The 15-ms window is determined by the 10-ms intersite propagation time plus a 5-ms allowance for uncertainty in the arrival time of weak signals. Both searches use a discrete bank of waveform templates [7,50–55] which target gravitational waves from binary black hole systems with a total mass of less than $100M_\odot$ and dimensionless spins aligned with the orbital angular momentum. Details of this bank are given in [18]. Identification by these two independent searches increases our confidence in the robustness and reliability of the detection.

The two searches employ different methods of ranking gravitational-wave candidates and techniques for estimating the noise background [14,17,18]. Each search defines a unique detection statistic to rank the likelihood of a candidate being a signal. The significance of a candidate event is estimated by comparing it with the noise background. This background is created using individual noise events produced in each detector's data. Since GW150914 had already been confirmed as a real gravitational-wave signal [4], it was removed from the data when estimating the noise background.

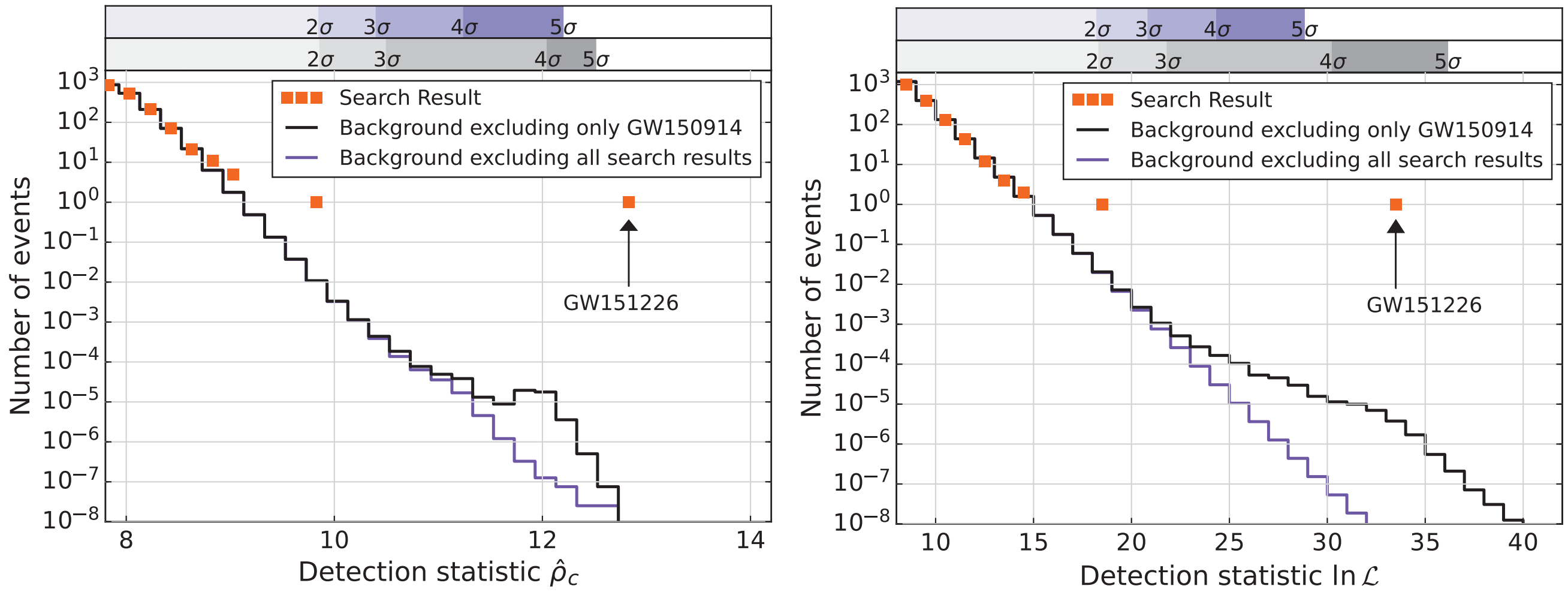

FIG. 2. Search results from the two binary coalescence searches using their respective detection statistics $\hat{\rho}_c$ (a combined matched filtering signal-to-noise ratio, defined precisely in [14]; left) and ln $\mathcal{L}$ (the log of a likelihood ratio, defined precisely in [17]; right). The event GW150914 is removed in all cases since it had already been confirmed as a real gravitational-wave signal. Both plots show the number of candidate events (search results) as a function of detection statistic with orange square markers. The mean number of background events as a function of the detection statistic is estimated using independent methods [18]. The background estimates are found using two methods: excluding all candidate events which are shown as orange square markers (purple lines) or including all candidate events except GW150914 (black lines). The scales along the top give the significance of an event in Gaussian standard deviations based on the corresponding noise background. The raised tail in the black-line background (left) is due to random coincidences of GW151226 in one detector with noise in the other detector and (right) due to the inclusion of GW151226 in the distribution of noise events in each detector. GW151226 is found with high significance in both searches. LVT151012 [5,18], visible in the search results at $\lesssim 2.0\sigma$, is the third most significant binary black hole candidate event in the observing period.

GW151226 was detected with a network matched-filter SNR of 13 by both searches. Figure 2 shows the detection statistic values assigned to GW151226 by the two searches and their respective noise background distributions. At the detection statistic value assigned to GW151226, the searches estimate a false alarm probability of $< 10^{-7}$ ($> 5\sigma$) [14] and $3.5 \times 10^{-6}$ ($4.5\sigma$) [17] when including candidate events in the background calculation. This procedure strictly limits the probability of obtaining a false positive outcome in the absence of signals [56]. The estimates from the two searches are consistent with expectations for a compact binary coalescence signal, given the differences in methods of data selection and candidate event ranking. When excluding search candidate events from the background calculation, a procedure that yields a mean-unbiased estimate of the distribution of noise events, the significance is found to be greater than $5\sigma$ in both searches. Further details of the noise background and significance estimation methods for each search are given in [18] and discussions specific to GW151226 are in [5].

## V. SOURCE DISCUSSION

To estimate the source parameters, a coherent Bayesian analysis [21,57] of the data was performed using two families of waveform models. Both models are calibrated to numerical simulations of binary black holes in general relativity. One waveform model includes spin-induced precession of the binary orbital plane [58], created by rotating the model described in [59]. The other waveform model restricts the component black hole spins to be aligned with the binary orbital angular momentum [42,43]. Both are publicly available [60]. Table I shows source parameters for GW151226 including the initial and final masses of the system. The parameter uncertainties include statistical and systematic errors from averaging posterior probability samples over the two waveform models, in addition to calibration uncertainties. Here, we report the median and 90% credible intervals.

The initial binary was composed of two stellar-mass black holes with a source-frame primary mass $m_1 = 14.2^{+8.3}_{-3.7} M_\odot$, secondary mass $m_2 = 7.5^{+2.3}_{-2.3} M_\odot$, and a total mass of $21.8^{+5.9}_{-1.7} M_\odot$. The binary merged into a black hole of mass $20.8^{+6.1}_{-1.7} M_\odot$, radiating $1.0^{+0.1}_{-0.2} M_\odot c^2$ in gravitational waves with a peak luminosity of $3.3^{+0.8}_{-1.6} \times 10^{56}$ erg/s. These estimates of the mass and spin of the final black hole, the total energy radiated in gravitational waves, and the peak gravitational-wave luminosity are derived from fits to numerical simulations [39,63–65]. The source localization is refined to 850 deg$^2$, owing to the different methods used [21], and refined calibration.

The long inspiral phase of GW151226 allows accurate estimates of lower-order post-Newtonian expansion parameters, such as the chirp mass [26,45]. However, only loose constraints can be placed on the total mass and mass ratio ($m_2/m_1$) because the merger and ringdown occur at frequencies where the detectors are less sensitive. Figure 3 shows the constraints on the component masses of the initial black hole binary. The component masses follow a line of constant chirp mass $8.9^{+0.3}_{-0.3} M_\odot$, and constrain the mass ratio to be greater than 0.28. The posterior distribution is not consistent with component masses below $4.5 M_\odot$ (99% credible level). This is above the theoretical maximum mass of a neutron star for common equations of state [66,67]. Thus, both components are identified as black holes.

TABLE I. Source parameters for GW151226. We report median values with 90% credible intervals that include statistical and systematic errors from averaging results of the precessing and nonprecessing spin waveform models. The errors also take into account calibration uncertainties. Masses are given in the source frame; to convert to the detector frame multiply by $(1 + z)$ [61]. The spins of the primary and secondary black holes are constrained to be positive. The source redshift assumes standard cosmology [62]. Further parameters of GW151226 are discussed in [5].

| | |
|---|---|
| Primary black hole mass | $14.2^{+8.3}_{-3.7} M_\odot$ |
| Secondary black hole mass | $7.5^{+2.3}_{-2.3} M_\odot$ |
| Chirp mass | $8.9^{+0.3}_{-0.3} M_\odot$ |
| Total black hole mass | $21.8^{+5.9}_{-1.7} M_\odot$ |
| Final black hole mass | $20.8^{+6.1}_{-1.7} M_\odot$ |
| Radiated gravitational-wave energy | $1.0^{+0.1}_{-0.2} M_\odot c^2$ |
| Peak luminosity | $3.3^{+0.8}_{-1.6} \times 10^{56}$ erg/s |
| Final black hole spin | $0.74^{+0.06}_{-0.06}$ |
| Luminosity distance | $440^{+180}_{-190}$ Mpc |
| Source redshift $z$ | $0.09^{+0.03}_{-0.04}$ |

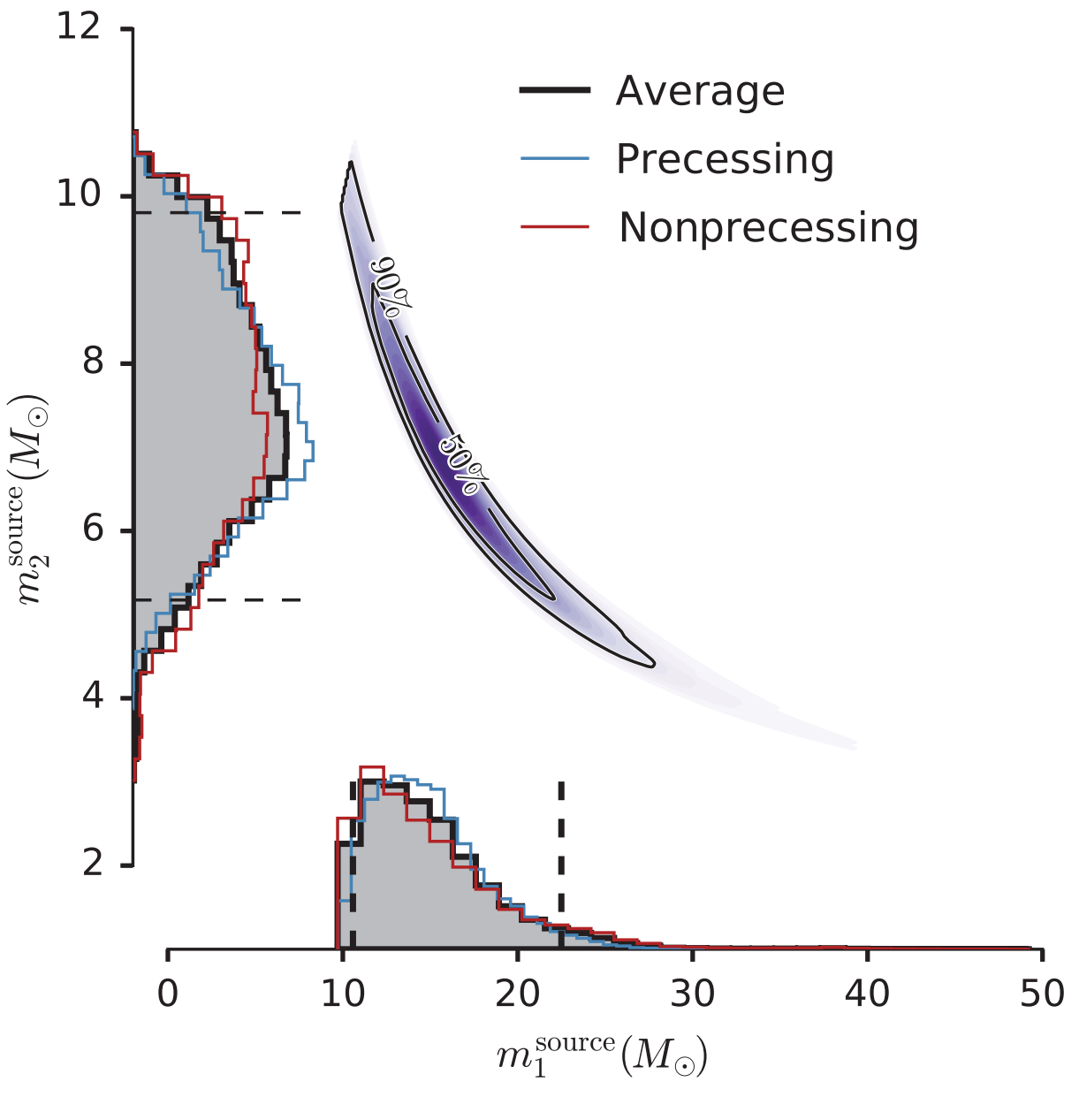

FIG. 3. Posterior density function for the source-frame masses $m_1^{\rm source}$ (primary) and $m_2^{\rm source}$ (secondary). The one-dimensional marginalized distributions include the posterior density functions for the precessing (blue) and nonprecessing (red) spin waveform models where average (black) represents the mean of the two models. The dashed lines mark the 90% credible interval for the average posterior density function. The two-dimensional plot shows the contours of the 50% and 90% credible regions plotted over a color-coded posterior density function.

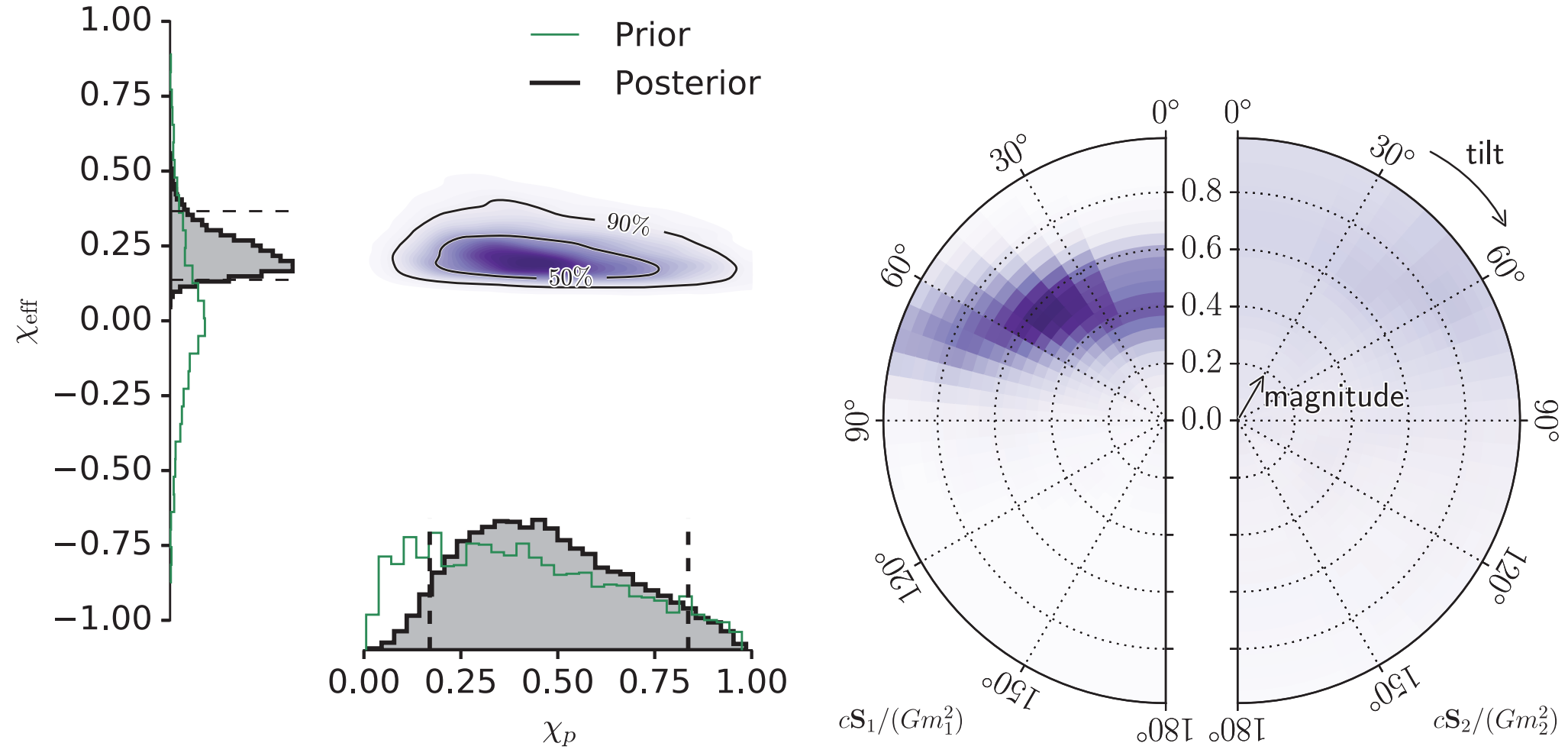

FIG. 4. *Left:* Posterior density function for the $\chi_p$ and $\chi_{\rm eff}$ spin parameters (measured at 20 Hz) compared to their prior distributions. The one-dimensional plot shows probability contours of the prior (green) and marginalized posterior density function (black) [58,59]. The two-dimensional plot shows the contours of the 50% and 90% credible regions plotted over a color-coded posterior density function. The dashed lines mark the 90% credible interval. *Right:* Posterior density function for the dimensionless component spins, $c\mathbf{S}_1/(Gm_1^2)$ and $c\mathbf{S}_2/(Gm_2^2)$, relative to the normal of the orbital plane $\hat{\mathbf{L}}$. $\mathbf{S}_i$ and $m_i$ are the spin angular momenta and masses of the primary ($i = 1$) and secondary ($i = 2$) black holes, $c$ is the speed of light and $G$ is the gravitational constant. The posterior density functions are marginalized over the azimuthal angles. The bins are designed to have equal prior probability; they are constructed linearly in spin magnitudes and the cosine of the tilt angles $\cos^{-1}(\hat{\mathbf{S}}_i \cdot \hat{\mathbf{L}})$.

Compact binary coalescences act as standard sirens [68,69]. Their luminosity distance can be extracted from the amplitude of an observed signal provided the orientation of the orbital plane can be determined. Information about whether the orbit is face-on, edge-on, or in between is encoded in the two polarizations of the gravitational wave. However, the two LIGO detectors are nearly coaligned and the source of GW151226 is likely to be located close to the maxima of the directional responses of both detectors [3]. Consequently, it is difficult to extract the polarization content, and therefore the orientation of the orbital plane. As a result, the luminosity distance is only weakly constrained to be $440^{+180}_{-190}$ Mpc, corresponding to a redshift of $0.09^{+0.03}_{-0.04}$ assuming a flat ΛCDM cosmology [62].

Component spins affect the relativistic motion of the binary but often have only subtle effects on the gravitational waveform. Therefore, we can only extract limited information about the spins. Figure 4 (left) shows the probability density functions of the mass-weighted combinations of orbit-aligned spins $\chi_{\rm eff}$ [70,71] and in-plane spins $\chi_p$ [72] for the precessing spin waveform model. The same figure (right) shows the individual spins of the component black holes. The posterior density functions inferred from the precessing and nonprecessing spin waveform models indicate that $\chi_{\rm eff}$ is positive at greater than the 99% credible level; therefore, at least one of the black holes has nonzero spin. We find that at least one black hole has a spin magnitude greater than 0.2 at the 99% credible level. Only weak constraints can be placed on $\chi_p$, suggesting that the data are not informative regarding spin-precession effects [5].

To test whether GW151226 is consistent with general relativity, we allow the coefficients that describe the waveform (which are derived as functions of the source parameters from the post-Newtonian approximation [26–28] and from fits to numerical relativity simulations) to deviate from their nominal values, and check whether the resulting waveforms are consistent with the data [73]. The posterior probability densities of the coefficients are found to center on their general relativity values. Additionally, both the offsets and widths of the posteriors for the post-Newtonian inspiral coefficients decrease significantly when analyzing GW150914 and GW151226 jointly, in some cases to the 10% level, as discussed in [5].

The waveform models used are consistent with general relativity simulations. Figure 5 shows GW151226's waveform reconstruction (90% credible region as in [57]) using the nonprecessing spin templates employed to find the signal and extract parameters, plotted during the time interval with the most significant SNR. Also shown is a direct numerical solution of Einstein's equations [38,74,75] for a binary black hole with parameters near the peak of the parameter estimation posterior.

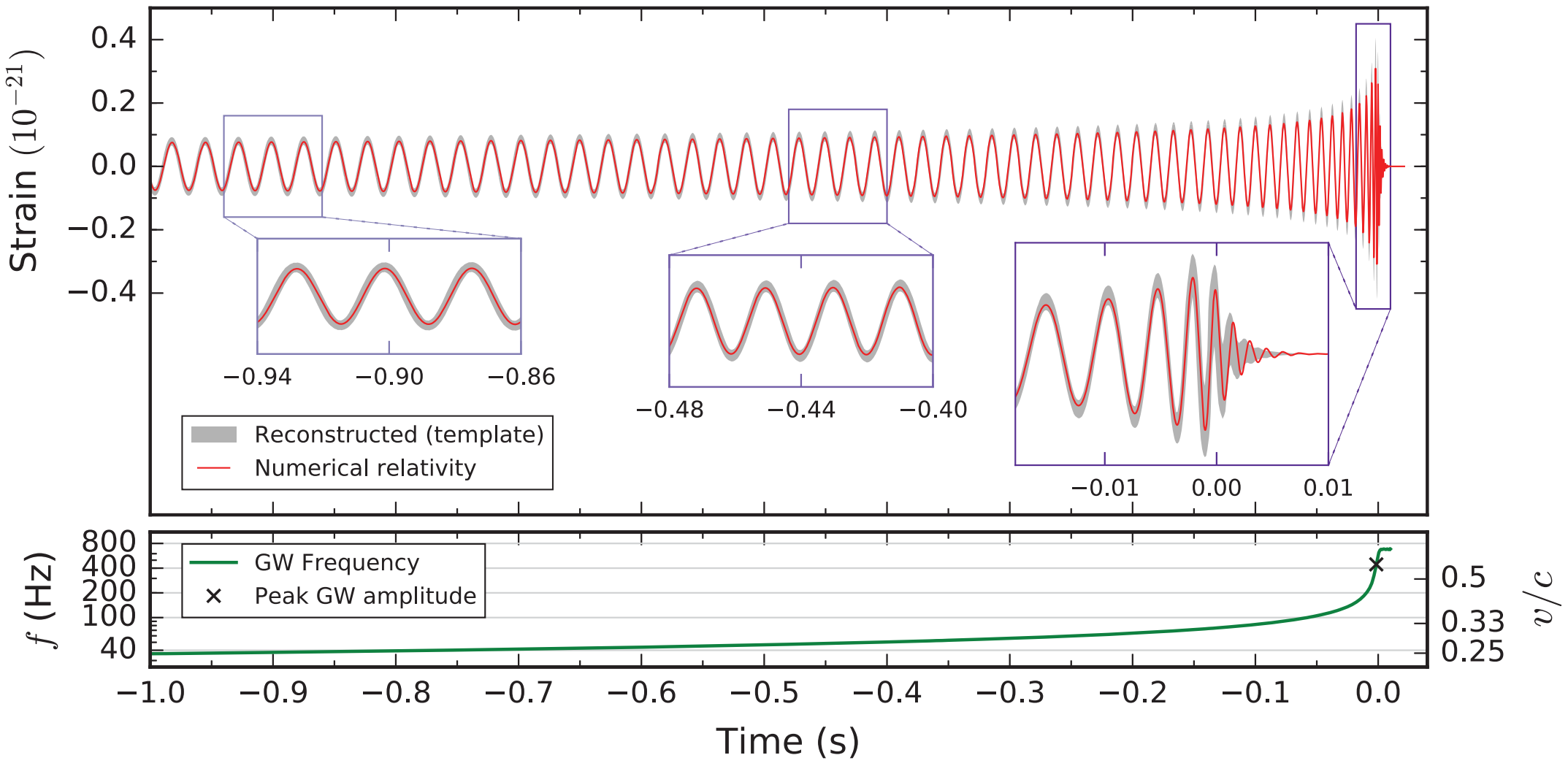

FIG. 5. Estimated gravitational-wave strain from GW151226 projected onto the LIGO Livingston detector with times relative to December 26, 2015 at 03:38:53.648 UTC. This shows the full bandwidth, without the filtering used for Fig. 1. *Top:* The 90% credible region (as in [57]) for a nonprecessing spin waveform-model reconstruction (gray) and a direct, nonprecessing numerical solution of Einstein's equations (red) with parameters consistent with the 90% credible region. *Bottom:* The gravitational-wave frequency $f$ (left axis) computed from the numerical-relativity waveform. The cross denotes the location of the maximum of the waveform amplitude, approximately coincident with the merger of the two black holes. During the inspiral, $f$ can be related to an effective relative velocity (right axis) given by the post-Newtonian parameter $v/c = (GM\pi f/c^3)^{1/3}$, where $M$ is the total mass.

## VI. ASTROPHYSICAL IMPLICATIONS

The inferred black hole masses are within the range of dynamically measured masses of black holes found in x-ray binaries [76–80], unlike GW150914. For the secondary black hole, there is a probability of 4% that it lies in the posited 3–$5M_\odot$ gap between observed neutron star and black hole masses [76,77], and there is no support for the primary black hole to have a mass in this range.

Binary black hole formation has been predicted through a range of different channels involving either isolated binaries or dynamical processes in dense stellar systems [81]. At present all types of formation channels predict binary black hole merger rates and black hole masses consistent with the observational constraints from GW150914 [82–84]. Both classical isolated binary evolution through the common envelope phase and dynamical formation are also consistent with GW151226, whose formation time and time delay to merger cannot be determined from the merger observation. Given our current understanding of massive-star evolution, the measured black hole masses are also consistent with any metallicity for the stellar progenitors and a broad range of progenitor masses [85,86].

The spin distribution of the black holes in stellar-mass binary black holes is unknown; the measurement of a spin magnitude for at least one companion greater than 0.2 is an important first step in constraining this distribution. Predictions of mass ratios and spin tilts with respect to the orbital angular momentum differ significantly for different channels. However, our current constraints on these properties are limited; implications for the evolutionary history of the observed black hole mergers are further discussed in [5].

The first observing period of Advanced LIGO provides evidence for a population of stellar-mass binary black holes contributing to a stochastic background that could be higher than previously expected [87]. Additionally, we find the rate estimate of stellar-mass binary black hole mergers in the local Universe to be consistent with the ranges presented in [88]. An updated discussion of the rate estimates can be found in [5].

A comprehensive discussion of inferred source parameters, astrophysical implications, mass distributions, rate estimations, and tests of general relativity for the binary black hole mergers detected during Advanced LIGO's first observing period may be found in [5].

## VII. CONCLUSION

LIGO has detected a second gravitational-wave signal from the coalescence of two stellar-mass black holes with lower masses than those measured for GW150914. Public data associated with GW151226 are available at [89]. The inferred component masses are consistent with values dynamically measured in x-ray binaries, but are obtained through the independent measurement process of gravitational-wave detection. Although it is challenging to constrain the spins of the initial black holes, we can conclude that at least one black hole had spin greater than 0.2. These recent detections in Advanced LIGO's first observing period have revealed a population of binary black holes that heralds the opening of the field of gravitational-wave astronomy.

## ACKNOWLEDGMENTS

The authors gratefully acknowledge the support of the United States National Science Foundation (NSF) for the construction and operation of the LIGO Laboratory and Advanced LIGO as well as the Science and Technology Facilities Council (STFC) of the United Kingdom, the Max Planck Society (MPS), and the State of Niedersachsen/Germany for support of the construction of Advanced LIGO and construction and operation of the GEO600 detector. Additional support for Advanced LIGO was provided by the Australian Research Council. The authors gratefully acknowledge the Italian Istituto Nazionale di Fisica Nucleare (INFN), the French Centre National de la Recherche Scientifique (CNRS) and the Foundation for Fundamental Research on Matter supported by the Netherlands Organisation for Scientific Research, for the construction and operation of the Virgo detector and the creation and support of the EGO consortium. The authors also gratefully acknowledge research support from these agencies as well as by the Council of Scientific and Industrial Research of India, Department of Science and Technology, India, Science & Engineering Research Board (SERB), India, Ministry of Human Resource Development, India, the Spanish Ministerio de Economía y Competitividad, the Conselleria d'Economia i Competitivitat and Conselleria d'Educació, Cultura i Universitats of the Govern de les Illes Balears, the National Science Centre of Poland, the European Commission, the Royal Society, the Scottish Funding Council, the Scottish Universities Physics Alliance, the Hungarian Scientific Research Fund (OTKA), the Lyon Institute of Origins (LIO), the National Research Foundation of Korea, Industry Canada and the Province of Ontario through the Ministry of Economic Development and Innovation, the Natural Sciences and Engineering Research Council of Canada, Canadian Institute for Advanced Research, the Brazilian Ministry of Science, Technology, and Innovation, Fundação de Amparo à Pesquisa do Estado de São Paulo (FAPESP), Russian Foundation for Basic Research, the Leverhulme Trust, the Research Corporation, Ministry of Science and Technology (MOST), Taiwan, and the Kavli Foundation. The authors gratefully acknowledge the support of the NSF, STFC, MPS, INFN, CNRS and the State of Niedersachsen/Germany for provision of computational resources.

This article has been assigned the document numbers LIGO-P151226 and VIR-0285A-16.

B. P. Abbott,[1] R. Abbott,[1] T. D. Abbott,[2] M. R. Abernathy,[3] F. Acernese,[4,5] K. Ackley,[6] C. Adams,[7] T. Adams,[8] P. Addesso,[9] R. X. Adhikari,[1] V. B. Adya,[10] C. Affeldt,[10] M. Agathos,[11] K. Agatsuma,[11] N. Aggarwal,[12] O. D. Aguiar,[13] L. Aiello,[14,15] A. Ain,[16] P. Ajith,[17] B. Allen,[10,18,19] A. Allocca,[20,21] P. A. Altin,[22] S. B. Anderson,[1] W. G. Anderson,[18] K. Arai,[1] M. C. Araya,[1] C. C. Arceneaux,[23] J. S. Areeda,[24] N. Arnaud,[25] K. G. Arun,[26] S. Ascenzi,[27,15] G. Ashton,[28] M. Ast,[29] S. M. Aston,[7] P. Astone,[30] P. Aufmuth,[19] C. Aulbert,[10] S. Babak,[31] P. Bacon,[32] M. K. M. Bader,[11] P. T. Baker,[33] F. Baldaccini,[34,35] G. Ballardin,[36] S. W. Ballmer,[37] J. C. Barayoga,[1] S. E. Barclay,[38] B. C. Barish,[1] D. Barker,[39] F. Barone,[4,5] B. Barr,[38] L. Barsotti,[12] M. Barsuglia,[32] D. Barta,[40] J. Bartlett,[39] I. Bartos,[41] R. Bassiri,[42] A. Basti,[20,21] J. C. Batch,[39] C. Baune,[10] V. Bavigadda,[36] M. Bazzan,[43,44] M. Bejger,[45] A. S. Bell,[38] B. K. Berger,[1] G. Bergmann,[10] C. P. L. Berry,[46] D. Bersanetti,[47,48] A. Bertolini,[11] J. Betzwieser,[7] S. Bhagwat,[37] R. Bhandare,[49] I. A. Bilenko,[50] G. Billingsley,[1] J. Birch,[7] R. Birney,[51] O. Birnholtz,[10] S. Biscans,[12] A. Bisht,[10,19] M. Bitossi,[36] C. Biwer,[37] M. A. Bizouard,[25] J. K. Blackburn,[1] C. D. Blair,[52] D. G. Blair,[52] R. M. Blair,[39] S. Bloemen,[53] O. Bock,[10] M. Boer,[54] G. Bogaert,[54] C. Bogan,[10] A. Bohe,[31] C. Bond,[46] F. Bondu,[55] R. Bonnand,[8] B. A. Boom,[11] R. Bork,[1] V. Boschi,[20,21] S. Bose,[56,16] Y. Bouffanais,[32] A. Bozzi,[36] C. Bradaschia,[21] P. R. Brady,[18] V. B. Braginsky,[50a] M. Branchesi,[57,58] J. E. Brau,[59] T. Briant,[60] A. Brillet,[54] M. Brinkmann,[10] V. Brisson,[25] P. Brockill,[18] J. E. Broida,[61] A. F. Brooks,[1] D. A. Brown,[37] D. D. Brown,[46] N. M. Brown,[12] S. Brunett,[1] C. C. Buchanan,[2] A. Buikema,[12] T. Bulik,[62] H. J. Bulten,[63,11] A. Buonanno,[31,64] D. Buskulic,[8] C. Buy,[32] R. L. Byer,[42] M. Cabero,[10] L. Cadonati,[65] G. Cagnoli,[66,67] C. Cahillane,[1] J. Calderón Bustillo,[65] T. Callister,[1] E. Calloni,[68,5] J. B. Camp,[69] K. C. Cannon,[70] J. Cao,[71] C. D. Capano,[10] E. Capocasa,[32] F. Carbognani,[36] S. Caride,[72] J. Casanueva Diaz,[25] C. Casentini,[27,15] S. Caudill,[18] M. Cavaglià,[23] F. Cavalier,[25] R. Cavalieri,[36] G. Cella,[21] C. B. Cepeda,[1] L. Cerboni Baiardi,[57,58] G. Cerretani,[20,21] E. Cesarini,[27,15] S. J. Chamberlin,[90] M. Chan,[38] S. Chao,[73] P. Charlton,[74] E. Chassande-Mottin,[32] B. D. Cheeseboro,[75] H. Y. Chen,[76] Y. Chen,[77] C. Cheng,[73] A. Chincarini,[48] A. Chiummo,[36] H. S. Cho,[78] M. Cho,[64] J. H. Chow,[22] N. Christensen,[61] Q. Chu,[52] S. Chua,[60] S. Chung,[52] G. Ciani,[6] F. Clara,[39] J. A. Clark,[65] F. Cleva,[54] E. Coccia,[27,14] P.-F. Cohadon,[60] A. Colla,[79,30] C. G. Collette,[80] L. Cominsky,[81] M. Constancio Jr.,[13] A. Conte,[79,30] L. Conti,[44] D. Cook,[39] T. R. Corbitt,[2] N. Cornish,[33] A. Corsi,[72] S. Cortese,[36] C. A. Costa,[13] M. W. Coughlin,[61] S. B. Coughlin,[82] J.-P. Coulon,[54] S. T. Countryman,[41] P. Couvares,[1] E. E. Cowan,[65] D. M. Coward,[52] M. J. Cowart,[7] D. C. Coyne,[1] R. Coyne,[72] K. Craig,[38] J. D. E. Creighton,[18] J. Cripe,[2] S. G. Crowder,[83] A. Cumming,[38] L. Cunningham,[38] E. Cuoco,[36] T. Dal Canton,[10] S. L. Danilishin,[38] S. D'Antonio,[15] K. Danzmann,[19,10] N. S. Darman,[84] A. Dasgupta,[85] C. F. Da Silva Costa,[6] V. Dattilo,[36] I. Dave,[49] M. Davier,[25] G. S. Davies,[38] E. J. Daw,[86] R. Day,[36] S. De,[37] D. DeBra,[42] G. Debreczeni,[40] J. Degallaix,[66] M. De Laurentis,[68,5] S. Deléglise,[60] W. Del Pozzo,[46] T. Denker,[10] T. Dent,[10] V. Dergachev,[1] R. De Rosa,[68,5] R. T. DeRosa,[7] R. DeSalvo,[9] R. C. Devine,[75] S. Dhurandhar,[16] M. C. Díaz,[87] L. Di Fiore,[5] M. Di Giovanni,[88,89] T. Di Girolamo,[68,5] A. Di Lieto,[20,21] S. Di Pace,[79,30] I. Di Palma,[31,79,30] A. Di Virgilio,[21] V. Dolique,[66] F. Donovan,[12] K. L. Dooley,[23] S. Doravari,[10] R. Douglas,[38] T. P. Downes,[18] M. Drago,[10] R. W. P. Drever,[1] J. C. Driggers,[39] M. Ducrot,[8] S. E. Dwyer,[39] T. B. Edo,[86] M. C. Edwards,[61] A. Effler,[7] H.-B. Eggenstein,[10] P. Ehrens,[1] J. Eichholz,[6,1] S. S. Eikenberry,[6] W. Engels,[77] R. C. Essick,[12] T. Etzel,[1] M. Evans,[12] T. M. Evans,[7] R. Everett,[90] M. Factourovich,[41] V. Fafone,[27,15] H. Fair,[37] S. Fairhurst,[91] X. Fan,[71] Q. Fang,[52] S. Farinon,[48] B. Farr,[76] W. M. Farr,[46] M. Favata,[92] M. Fays,[91] H. Fehrmann,[10] M. M. Fejer,[42] E. Fenyvesi,[93] I. Ferrante,[20,21] E. C. Ferreira,[13] F. Ferrini,[36] F. Fidecaro,[20,21] I. Fiori,[36] D. Fiorucci,[32] R. P. Fisher,[37] R. Flaminio,[66,94] M. Fletcher,[38] H. Fong,[95] J.-D. Fournier,[54] S. Frasca,[79,30] F. Frasconi,[21] Z. Frei,[93] A. Freise,[46] R. Frey,[59] V. Frey,[25] P. Fritschel,[12] V. V. Frolov,[7] P. Fulda,[6] M. Fyffe,[7] H. A. G. Gabbard,[23] J. R. Gair,[96] L. Gammaitoni,[34] S. G. Gaonkar,[16] F. Garufi,[68,5] G. Gaur,[97,85] N. Gehrels,[69] G. Gemme,[48] P. Geng,[87] E. Genin,[36] A. Gennai,[21] J. George,[49] L. Gergely,[98] V. Germain,[8] Abhirup Ghosh,[17] Archisman Ghosh,[17] S. Ghosh,[53,11] J. A. Giaime,[2,7] K. D. Giardina,[7] A. Giazotto,[21] K. Gill,[99] A. Glaefke,[38] E. Goetz,[39] R. Goetz,[6] L. Gondan,[93] G. González,[2] J. M. Gonzalez Castro,[20,21] A. Gopakumar,[100] N. A. Gordon,[38] M. L. Gorodetsky,[50] S. E. Gossan,[1] M. Gosselin,[36] R. Gouaty,[8] A. Grado,[101,5] C. Graef,[38] P. B. Graff,[64] M. Granata,[66] A. Grant,[38] S. Gras,[12] C. Gray,[39] G. Greco,[57,58]

A. C. Green,[46] P. Groot,[53] H. Grote,[10] S. Grunewald,[31] G. M. Guidi,[57,58] X. Guo,[71] A. Gupta,[16] M. K. Gupta,[85] K. E. Gushwa,[1] E. K. Gustafson,[1] R. Gustafson,[102] J. J. Hacker,[24] B. R. Hall,[56] E. D. Hall,[1] H. Hamilton,[103] G. Hammond,[38] M. Haney,[100] M. M. Hanke,[10] J. Hanks,[39] C. Hanna,[90] M. D. Hannam,[91] J. Hanson,[7] T. Hardwick,[2] J. Harms,[57,58] G. M. Harry,[3] I. W. Harry,[31] M. J. Hart,[38] M. T. Hartman,[6] C.-J. Haster,[46] K. Haughian,[38] J. Healy,[104] A. Heidmann,[60] M. C. Heintze,[7] H. Heitmann,[54] P. Hello,[25] G. Hemming,[36] M. Hendry,[38] I. S. Heng,[38] J. Hennig,[38] J. Henry,[104] A. W. Heptonstall,[1] M. Heurs,[10,19] S. Hild,[38] D. Hoak,[36] D. Hofman,[66] K. Holt,[7] D. E. Holz,[76] P. Hopkins,[91] J. Hough,[38] E. A. Houston,[38] E. J. Howell,[52] Y. M. Hu,[10] S. Huang,[73] E. A. Huerta,[105] D. Huet,[25] B. Hughey,[99] S. Husa,[106] S. H. Huttner,[38] T. Huynh-Dinh,[7] N. Indik,[10] D. R. Ingram,[39] R. Inta,[72] H. N. Isa,[38] J.-M. Isac,[60] M. Isi,[1] T. Isogai,[12] B. R. Iyer,[17] K. Izumi,[39] T. Jacqmin,[60] H. Jang,[78] K. Jani,[65] P. Jaranowski,[107] S. Jawahar,[108] L. Jian,[52] F. Jiménez-Forteza,[106] W. W. Johnson,[2] N. K. Johnson-McDaniel,[17] D. I. Jones,[28] R. Jones,[38] R. J. G. Jonker,[11] L. Ju,[52] Haris K,[109] C. V. Kalaghatgi,[91] V. Kalogera,[82] S. Kandhasamy,[23] G. Kang,[78] J. B. Kanner,[1] S. J. Kapadia,[10] S. Karki,[59] K. S. Karvinen,[10] M. Kasprzack,[36,2] E. Katsavounidis,[12] W. Katzman,[7] S. Kaufer,[19] T. Kaur,[52] K. Kawabe,[39] F. Kéfélian,[54] M. S. Kehl,[95] D. Keitel,[106] D. B. Kelley,[37] W. Kells,[1] R. Kennedy,[86] J. S. Key,[87] F. Y. Khalili,[50] I. Khan,[14] S. Khan,[91] Z. Khan,[85] E. A. Khazanov,[110] N. Kijbunchoo,[39] Chi-Woong Kim,[78] Chunglee Kim,[78] J. Kim,[111] K. Kim,[112] N. Kim,[42] W. Kim,[113] Y.-M. Kim,[111] S. J. Kimbrell,[65] E. J. King,[113] P. J. King,[39] J. S. Kissel,[39] B. Klein,[82] L. Kleybolte,[29] S. Klimenko,[6] S. M. Koehlenbeck,[10] S. Koley,[11] V. Kondrashov,[1] A. Kontos,[12] M. Korobko,[29] W. Z. Korth,[1] I. Kowalska,[62] D. B. Kozak,[1] V. Kringel,[10] B. Krishnan,[10] A. Królak,[114,115] C. Krueger,[19] G. Kuehn,[10] P. Kumar,[95] R. Kumar,[85] L. Kuo,[73] A. Kutynia,[114] B. D. Lackey,[37] M. Landry,[39] J. Lange,[104] B. Lantz,[42] P. D. Lasky,[116] M. Laxen,[7] A. Lazzarini,[1] C. Lazzaro,[44] P. Leaci,[79,30] S. Leavey,[38] E. O. Lebigot,[32,71] C. H. Lee,[111] H. K. Lee,[112] H. M. Lee,[117] K. Lee,[38] A. Lenon,[37] M. Leonardi,[88,89] J. R. Leong,[10] N. Leroy,[25] N. Letendre,[8] Y. Levin,[116] J. B. Lewis,[1] T. G. F. Li,[118] A. Libson,[12] T. B. Littenberg,[119] N. A. Lockerbie,[108] A. L. Lombardi,[120] L. T. London,[91] J. E. Lord,[37] M. Lorenzini,[14,15] V. Loriette,[121] M. Lormand,[7] G. Losurdo,[58] J. D. Lough,[10,19] C. O. Lousto,[104] H. Lück,[19,10] A. P. Lundgren,[10] R. Lynch,[12] Y. Ma,[52] B. Machenschalk,[10] M. MacInnis,[12] D. M. Macleod,[2] F. Magaña-Sandoval,[37] L. Magaña Zertuche,[37] R. M. Magee,[56] E. Majorana,[30] I. Maksimovic,[121] V. Malvezzi,[27,15] N. Man,[54] I. Mandel,[46] V. Mandic,[83] V. Mangano,[38] G. L. Mansell,[22] M. Manske,[18] M. Mantovani,[36] F. Marchesoni,[122,35] F. Marion,[8] S. Márka,[41] Z. Márka,[41] A. S. Markosyan,[42] E. Maros,[1] F. Martelli,[57,58] L. Martellini,[54] I. W. Martin,[38] D. V. Martynov,[12] J. N. Marx,[1] K. Mason,[12] A. Masserot,[8] T. J. Massinger,[37] M. Masso-Reid,[38] S. Mastrogiovanni,[79,30] F. Matichard,[12] L. Matone,[41] N. Mavalvala,[12] N. Mazumder,[56] R. McCarthy,[39] D. E. McClelland,[22] S. McCormick,[7] S. C. McGuire,[123] G. McIntyre,[1] J. McIver,[1] D. J. McManus,[22] T. McRae,[22] S. T. McWilliams,[75] D. Meacher,[90] G. D. Meadors,[31,10] J. Meidam,[11] A. Melatos,[84] G. Mendell,[39] R. A. Mercer,[18] E. L. Merilh,[39] M. Merzougui,[54] S. Meshkov,[1] C. Messenger,[38] C. Messick,[90] R. Metzdorff,[60] P. M. Meyers,[83] F. Mezzani,[30,79] H. Miao,[46] C. Michel,[66] H. Middleton,[46] E. E. Mikhailov,[124] L. Milano,[68,5] A. L. Miller,[6,79,30] A. Miller,[82] B. B. Miller,[82] J. Miller,[12] M. Millhouse,[33] Y. Minenkov,[15] J. Ming,[31] S. Mirshekari,[125] C. Mishra,[17] S. Mitra,[16] V. P. Mitrofanov,[50] G. Mitselmakher,[6] R. Mittleman,[12] A. Moggi,[21] M. Mohan,[36] S. R. P. Mohapatra,[12] M. Montani,[57,58] B. C. Moore,[92] C. J. Moore,[126] D. Moraru,[39] G. Moreno,[39] S. R. Morriss,[87] K. Mossavi,[10] B. Mours,[8] C. M. Mow-Lowry,[46] G. Mueller,[6] A. W. Muir,[91] Arunava Mukherjee,[17] D. Mukherjee,[18] S. Mukherjee,[87] N. Mukund,[16] A. Mullavey,[7] J. Munch,[113] D. J. Murphy,[41] P. G. Murray,[38] A. Mytidis,[6] I. Nardecchia,[27,15] L. Naticchioni,[79,30] R. K. Nayak,[127] K. Nedkova,[120] G. Nelemans,[53,11] T. J. N. Nelson,[7] M. Neri,[47,48] A. Neunzert,[102] G. Newton,[38] T. T. Nguyen,[22] A. B. Nielsen,[10] S. Nissanke,[53,11] A. Nitz,[10] F. Nocera,[36] D. Nolting,[7] M. E. N. Normandin,[87] L. K. Nuttall,[37] J. Oberling,[39] E. Ochsner,[18] J. O'Dell,[128] E. Oelker,[12] G. H. Ogin,[129] J. J. Oh,[130] S. H. Oh,[130] F. Ohme,[91] M. Oliver,[106] P. Oppermann,[10] Richard J. Oram,[7] B. O'Reilly,[7] R. O'Shaughnessy,[104] D. J. Ottaway,[113] H. Overmier,[7] B. J. Owen,[72] A. Pai,[109] S. A. Pai,[49] J. R. Palamos,[59] O. Palashov,[110] C. Palomba,[30] A. Pal-Singh,[29] H. Pan,[73] C. Pankow,[82] F. Pannarale,[91] B. C. Pant,[49] F. Paoletti,[36,21] A. Paoli,[36] M. A. Papa,[31,18,10] H. R. Paris,[42] W. Parker,[7] D. Pascucci,[38] A. Pasqualetti,[36] R. Passaquieti,[20,21] D. Passuello,[21] B. Patricelli,[20,21] Z. Patrick,[42] B. L. Pearlstone,[38] M. Pedraza,[1] R. Pedurand,[66,131] L. Pekowsky,[37] A. Pele,[7] S. Penn,[132] A. Perreca,[1] L. M. Perri,[82] H. P. Pfeiffer,[95,31] M. Phelps,[38] O. J. Piccinni,[79,30] M. Pichot,[54] F. Piergiovanni,[57,58] V. Pierro,[9] G. Pillant,[36] L. Pinard,[66] I. M. Pinto,[9] M. Pitkin,[38] M. Poe,[18] R. Poggiani,[20,21] P. Popolizio,[36] A. Post,[10] J. Powell,[38] J. Prasad,[16] V. Predoi,[91] T. Prestegard,[83] L. R. Price,[1] M. Prijatelj,[10,36] M. Principe,[9] S. Privitera,[31] R. Prix,[10] G. A. Prodi,[88,89] L. Prokhorov,[50] O. Puncken,[10] M. Punturo,[35] P. Puppo,[30] M. Pürrer,[31] H. Qi,[18] J. Qin,[52] S. Qiu,[116] V. Quetschke,[87] E. A. Quintero,[1] R. Quitzow-James,[59] F. J. Raab,[39] D. S. Rabeling,[22] H. Radkins,[39] P. Raffai,[93] S. Raja,[49] C. Rajan,[49] M. Rakhmanov,[87] P. Rapagnani,[79,30] V. Raymond,[31] M. Razzano,[20,21] V. Re,[27] J. Read,[24] C. M. Reed,[39]

T. Regimbau,[54] L. Rei,[48] S. Reid,[51] D. H. Reitze,[1,6] H. Rew,[124] S. D. Reyes,[37] F. Ricci,[79,30] K. Riles,[102] M. Rizzo,[104] N. A. Robertson,[1,38] R. Robie,[38] F. Robinet,[25] A. Rocchi,[15] L. Rolland,[8] J. G. Rollins,[1] V. J. Roma,[59] J. D. Romano,[87] R. Romano,[4,5] G. Romanov,[124] J. H. Romie,[7] D. Rosińska,[133,45] S. Rowan,[38] A. Rüdiger,[10] P. Ruggi,[36] K. Ryan,[39] S. Sachdev,[1] T. Sadecki,[39] L. Sadeghian,[18] M. Sakellariadou,[134] L. Salconi,[36] M. Saleem,[109] F. Salemi,[10] A. Samajdar,[127] L. Sammut,[116] E. J. Sanchez,[1] V. Sandberg,[39] B. Sandeen,[82] J. R. Sanders,[37] B. Sassolas,[66] B. S. Sathyaprakash,[91] P. R. Saulson,[37] O. E. S. Sauter,[102] R. L. Savage,[39] A. Sawadsky,[19] P. Schale,[59] R. Schilling,[10b] J. Schmidt,[10] P. Schmidt,[1,77] R. Schnabel,[29] R. M. S. Schofield,[59] A. Schönbeck,[29] E. Schreiber,[10] D. Schuette,[10,19] B. F. Schutz,[91,31] J. Scott,[38] S. M. Scott,[22] D. Sellers,[7] A. S. Sengupta,[97] D. Sentenac,[36] V. Sequino,[27,15] A. Sergeev,[110] Y. Setyawati,[53,11] D. A. Shaddock,[22] T. Shaffer,[39] M. S. Shahriar,[82] M. Shaltev,[10] B. Shapiro,[42] P. Shawhan,[64] A. Sheperd,[18] D. H. Shoemaker,[12] D. M. Shoemaker,[65] K. Siellez,[65] X. Siemens,[18] M. Sieniawska,[45] D. Sigg,[39] A. D. Silva,[13] A. Singer,[1] L. P. Singer,[69] A. Singh,[31,10,19] R. Singh,[2] A. Singhal,[14] A. M. Sintes,[106] B. J. J. Slagmolen,[22] J. R. Smith,[24] N. D. Smith,[1] R. J. E. Smith,[1] E. J. Son,[130] B. Sorazu,[38] F. Sorrentino,[48] T. Souradeep,[16] A. K. Srivastava,[85] A. Staley,[41] M. Steinke,[10] J. Steinlechner,[38] S. Steinlechner,[38] D. Steinmeyer,[10,19] B. C. Stephens,[18] S. P. Stevenson,[46] R. Stone,[87] K. A. Strain,[38] N. Straniero,[66] G. Stratta,[57,58] N. A. Strauss,[61] S. Strigin,[50] R. Sturani,[125] A. L. Stuver,[7] T. Z. Summerscales,[135] L. Sun,[84] S. Sunil,[85] P. J. Sutton,[91] B. L. Swinkels,[36] M. J. Szczepańczyk,[99] M. Tacca,[32] D. Talukder,[59] D. B. Tanner,[6] M. Tápai,[98] S. P. Tarabrin,[10] A. Taracchini,[31] R. Taylor,[1] T. Theeg,[10] M. P. Thirugnanasambandam,[1] E. G. Thomas,[46] M. Thomas,[7] P. Thomas,[39] K. A. Thorne,[7] E. Thrane,[116] S. Tiwari,[14,89] V. Tiwari,[91] K. V. Tokmakov,[108] K. Toland,[38] C. Tomlinson,[86] M. Tonelli,[20,21] Z. Tornasi,[38] C. V. Torres,[87c] C. I. Torrie,[1] D. Töyrä,[46] F. Travasso,[34,35] G. Traylor,[7] D. Trifirò,[23] M. C. Tringali,[88,89] L. Trozzo,[136,21] M. Tse,[12] M. Turconi,[54] D. Tuyenbayev,[87] D. Ugolini,[137] C. S. Unnikrishnan,[100] A. L. Urban,[18] S. A. Usman,[37] H. Vahlbruch,[19] G. Vajente,[1] G. Valdes,[87] M. Vallisneri,[77] N. van Bakel,[11] M. van Beuzekom,[11] J. F. J. van den Brand,[63,11] C. Van Den Broeck,[11] D. C. Vander-Hyde,[37] L. van der Schaaf,[11] J. V. van Heijningen,[11] A. A. van Veggel,[38] M. Vardaro,[43,44] S. Vass,[1] M. Vasúth,[40] R. Vaulin,[12] A. Vecchio,[46] G. Vedovato,[44] J. Veitch,[46] P. J. Veitch,[113] K. Venkateswara,[138] D. Verkindt,[8] F. Vetrano,[57,58] A. Viceré,[57,58] S. Vinciguerra,[46] D. J. Vine,[51] J.-Y. Vinet,[54] S. Vitale,[12] T. Vo,[37] H. Vocca,[34,35] C. Vorvick,[39] D. V. Voss,[6] W. D. Vousden,[46] S. P. Vyatchanin,[50] A. R. Wade,[22] L. E. Wade,[139] M. Wade,[139] M. Walker,[2] L. Wallace,[1] S. Walsh,[31,10] G. Wang,[14,58] H. Wang,[46] M. Wang,[46] X. Wang,[71] Y. Wang,[52] R. L. Ward,[22] J. Warner,[39] M. Was,[8] B. Weaver,[39] L.-W. Wei,[54] M. Weinert,[10] A. J. Weinstein,[1] R. Weiss,[12] L. Wen,[52] P. Weßels,[10] T. Westphal,[10] K. Wette,[10] J. T. Whelan,[104] B. F. Whiting,[6] R. D. Williams,[1] A. R. Williamson,[91] J. L. Willis,[103] B. Willke,[19,10] M. H. Wimmer,[10,19] W. Winkler,[10] C. C. Wipf,[1] H. Wittel,[10,19] G. Woan,[38] J. Woehler,[10] J. Worden,[39] J. L. Wright,[38] D. S. Wu,[10] G. Wu,[7] J. Yablon,[82] W. Yam,[12] H. Yamamoto,[1] C. C. Yancey,[64] H. Yu,[12] M. Yvert,[8] A. Zadrożny,[114] L. Zangrando,[44] M. Zanolin,[99] J.-P. Zendri,[44] M. Zevin,[82] L. Zhang,[1] M. Zhang,[124] Y. Zhang,[104] C. Zhao,[52] M. Zhou,[82] Z. Zhou,[82] X. J. Zhu,[52] M. E. Zucker,[1,12] S. E. Zuraw,[120] and J. Zweizig[1]

(LIGO Scientific Collaboration and Virgo Collaboration)

M. Boyle,[140] D. Hemberger,[77] L. E. Kidder,[140] G. Lovelace,[24] S. Ossokine,[31] M. Scheel,[77] B. Szilagyi,[77,141] and S. Teukolsky[140]

[1]*LIGO, California Institute of Technology, Pasadena, California 91125, USA*

[2]*Louisiana State University, Baton Rouge, Louisiana 70803, USA*

[3]*American University, Washington, DC 20016, USA*

[4]*Università di Salerno, Fisciano, I-84084 Salerno, Italy*

[5]*INFN, Sezione di Napoli, Complesso Universitario di Monte S.Angelo, I-80126 Napoli, Italy*

[6]*University of Florida, Gainesville, Florida 32611, USA*

[7]*LIGO Livingston Observatory, Livingston, Louisiana 70754, USA*

[8]*Laboratoire d'Annecy-le-Vieux de Physique des Particules (LAPP), Université Savoie Mont Blanc, CNRS/IN2P3, F-74941 Annecy-le-Vieux, France*

[9]*University of Sannio at Benevento, I-82100 Benevento, Italy and INFN, Sezione di Napoli, I-80100 Napoli, Italy*

[10]*Albert-Einstein-Institut, Max-Planck-Institut für Gravitationsphysik, D-30167 Hannover, Germany*

[11]*Nikhef, Science Park, 1098 XG Amsterdam, Netherlands*

[12]*LIGO, Massachusetts Institute of Technology, Cambridge, Massachusetts 02139, USA*

[13]*Instituto Nacional de Pesquisas Espaciais, 12227-010 São José dos Campos, São Paulo, Brazil*

[14]INFN, Gran Sasso Science Institute, I-67100 L'Aquila, Italy
[15]INFN, Sezione di Roma Tor Vergata, I-00133 Roma, Italy
[16]Inter-University Centre for Astronomy and Astrophysics, Pune 411007, India
[17]International Centre for Theoretical Sciences, Tata Institute of Fundamental Research, Bangalore 560012, India
[18]University of Wisconsin-Milwaukee, Milwaukee, Wisconsin 53201, USA
[19]Leibniz Universität Hannover, D-30167 Hannover, Germany
[20]Università di Pisa, I-56127 Pisa, Italy
[21]INFN, Sezione di Pisa, I-56127 Pisa, Italy
[22]Australian National University, Canberra, Australian Capital Territory 0200, Australia
[23]The University of Mississippi, University, Mississippi 38677, USA
[24]California State University Fullerton, Fullerton, California 92831, USA
[25]LAL, Univ. Paris-Sud, CNRS/IN2P3, Université Paris-Saclay, Orsay, France
[26]Chennai Mathematical Institute, Chennai 603103, India
[27]Università di Roma Tor Vergata, I-00133 Roma, Italy
[28]University of Southampton, Southampton SO17 1BJ, United Kingdom
[29]Universität Hamburg, D-22761 Hamburg, Germany
[30]INFN, Sezione di Roma, I-00185 Roma, Italy
[31]Albert-Einstein-Institut, Max-Planck-Institut für Gravitationsphysik, D-14476 Potsdam-Golm, Germany
[32]APC, AstroParticule et Cosmologie, Université Paris Diderot, CNRS/IN2P3, CEA/Irfu, Observatoire de Paris, Sorbonne Paris Cité, F-75205 Paris Cedex 13, France
[33]Montana State University, Bozeman, Montana 59717, USA
[34]Università di Perugia, I-06123 Perugia, Italy
[35]INFN, Sezione di Perugia, I-06123 Perugia, Italy
[36]European Gravitational Observatory (EGO), I-56021 Cascina, Pisa, Italy
[37]Syracuse University, Syracuse, New York 13244, USA
[38]SUPA, University of Glasgow, Glasgow G12 8QQ, United Kingdom
[39]LIGO Hanford Observatory, Richland, Washington 99352, USA
[40]Wigner RCP, RMKI, H-1121 Budapest, Konkoly Thege Miklós út 29-33, Hungary
[41]Columbia University, New York, New York 10027, USA
[42]Stanford University, Stanford, California 94305, USA
[43]Università di Padova, Dipartimento di Fisica e Astronomia, I-35131 Padova, Italy
[44]INFN, Sezione di Padova, I-35131 Padova, Italy
[45]CAMK-PAN, 00-716 Warsaw, Poland
[46]University of Birmingham, Birmingham B15 2TT, United Kingdom
[47]Università degli Studi di Genova, I-16146 Genova, Italy
[48]INFN, Sezione di Genova, I-16146 Genova, Italy
[49]RRCAT, Indore MP 452013, India
[50]Faculty of Physics, Lomonosov Moscow State University, Moscow 119991, Russia
[51]SUPA, University of the West of Scotland, Paisley PA1 2BE, United Kingdom
[52]University of Western Australia, Crawley, Western Australia 6009, Australia
[53]Department of Astrophysics/IMAPP, Radboud University Nijmegen, P.O. Box 9010, 6500 GL Nijmegen, The Netherlands
[54]Artemis, Université Côte d'Azur, CNRS, Observatoire Côte d'Azur, CS 34229, Nice cedex 4, France
[55]Institut de Physique de Rennes, CNRS, Université de Rennes 1, F-35042 Rennes, France
[56]Washington State University, Pullman, Washington 99164, USA
[57]Università degli Studi di Urbino "Carlo Bo", I-61029 Urbino, Italy
[58]INFN, Sezione di Firenze, I-50019 Sesto Fiorentino, Firenze, Italy
[59]University of Oregon, Eugene, Oregon 97403, USA
[60]Laboratoire Kastler Brossel, UPMC-Sorbonne Universités, CNRS, ENS-PSL Research University, Collège de France, F-75005 Paris, France
[61]Carleton College, Northfield, Minnesota 55057, USA
[62]Astronomical Observatory Warsaw University, 00-478 Warsaw, Poland
[63]VU University Amsterdam, 1081 HV Amsterdam, Netherlands
[64]University of Maryland, College Park, Maryland 20742, USA
[65]Center for Relativistic Astrophysics and School of Physics, Georgia Institute of Technology, Atlanta, Georgia 30332, USA
[66]Laboratoire des Matériaux Avancés (LMA), CNRS/IN2P3, F-69622 Villeurbanne, France
[67]Université Claude Bernard Lyon 1, F-69622 Villeurbanne, France
[68]Università di Napoli "Federico II", Complesso Universitario di Monte S.Angelo, I-80126 Napoli, Italy
[69]NASA/Goddard Space Flight Center, Greenbelt, Maryland 20771, USA
[70]RESCEU, University of Tokyo, Tokyo, 113-0033, Japan
[71]Tsinghua University, Beijing 100084, China

[72]Texas Tech University, Lubbock, Texas 79409, USA
[73]National Tsing Hua University, Hsinchu City, 30013 Taiwan, Republic of China
[74]Charles Sturt University, Wagga Wagga, New South Wales 2678, Australia
[75]West Virginia University, Morgantown, West Virginia 26506, USA
[76]University of Chicago, Chicago, Illinois 60637, USA
[77]Caltech CaRT, Pasadena, California 91125, USA
[78]Korea Institute of Science and Technology Information, Daejeon 305-806, Korea
[79]Università di Roma "La Sapienza", I-00185 Roma, Italy
[80]University of Brussels, Brussels 1050, Belgium
[81]Sonoma State University, Rohnert Park, California 94928, USA
[82]Center for Interdisciplinary Exploration & Research in Astrophysics (CIERA), Northwestern University, Evanston, Illinois 60208, USA
[83]University of Minnesota, Minneapolis, Minnesota 55455, USA
[84]The University of Melbourne, Parkville, Victoria 3010, Australia
[85]Institute for Plasma Research, Bhat, Gandhinagar 382428, India
[86]The University of Sheffield, Sheffield S10 2TN, United Kingdom
[87]The University of Texas Rio Grande Valley, Brownsville, TX 78520, USA
[88]Università di Trento, Dipartimento di Fisica, I-38123 Povo, Trento, Italy
[89]INFN, Trento Institute for Fundamental Physics and Applications, I-38123 Povo, Trento, Italy
[90]The Pennsylvania State University, University Park, Pennsylvania 16802, USA
[91]Cardiff University, Cardiff CF24 3AA, United Kingdom
[92]Montclair State University, Montclair, New Jersey 07043, USA
[93]MTA Eötvös University, "Lendulet" Astrophysics Research Group, Budapest 1117, Hungary
[94]National Astronomical Observatory of Japan, 2-21-1 Osawa, Mitaka, Tokyo 181-8588, Japan
[95]Canadian Institute for Theoretical Astrophysics, University of Toronto, Toronto, Ontario M5S 3H8, Canada
[96]School of Mathematics, University of Edinburgh, Edinburgh EH9 3FD, United Kingdom
[97]Indian Institute of Technology, Gandhinagar Ahmedabad Gujarat 382424, India
[98]University of Szeged, Dóm tér 9, Szeged 6720, Hungary
[99]Embry-Riddle Aeronautical University, Prescott, Arizona 86301, USA
[100]Tata Institute of Fundamental Research, Mumbai 400005, India
[101]INAF, Osservatorio Astronomico di Capodimonte, I-80131, Napoli, Italy
[102]University of Michigan, Ann Arbor, Michigan 48109, USA
[103]Abilene Christian University, Abilene, Texas 79699, USA
[104]Rochester Institute of Technology, Rochester, New York 14623, USA
[105]NCSA, University of Illinois at Urbana-Champaign, Urbana, Illinois 61801, USA
[106]Universitat de les Illes Balears, IAC3—IEEC, E-07122 Palma de Mallorca, Spain
[107]University of Białystok, 15-424 Białystok, Poland
[108]SUPA, University of Strathclyde, Glasgow G1 1XQ, United Kingdom
[109]IISER-TVM, CET Campus, Trivandrum Kerala 695016, India
[110]Institute of Applied Physics, Nizhny Novgorod, 603950, Russia
[111]Pusan National University, Busan 609-735, Korea
[112]Hanyang University, Seoul 133-791, Korea
[113]University of Adelaide, Adelaide, South Australia 5005, Australia
[114]NCBJ, 05-400 Świerk-Otwock, Poland
[115]IM-PAN, 00-956 Warsaw, Poland
[116]Monash University, Victoria 3800, Australia
[117]Seoul National University, Seoul 151-742, Korea
[118]The Chinese University of Hong Kong, Shatin, NT, Hong Kong SAR, China
[119]University of Alabama in Huntsville, Huntsville, Alabama 35899, USA
[120]University of Massachusetts-Amherst, Amherst, Massachusetts 01003, USA
[121]ESPCI, CNRS, F-75005 Paris, France
[122]Università di Camerino, Dipartimento di Fisica, I-62032 Camerino, Italy
[123]Southern University and A&M College, Baton Rouge, Louisiana 70813, USA
[124]College of William and Mary, Williamsburg, Virginia 23187, USA
[125]Instituto de Física Teórica, University Estadual Paulista/ICTP South American Institute for Fundamental Research, São Paulo SP 01140-070, Brazil
[126]University of Cambridge, Cambridge CB2 1TN, United Kingdom
[127]IISER-Kolkata, Mohanpur, West Bengal 741252, India
[128]Rutherford Appleton Laboratory, HSIC, Chilton, Didcot, Oxon OX11 0QX, United Kingdom
[129]Whitman College, 345 Boyer Avenue, Walla Walla, Washington 99362 USA

[130]*National Institute for Mathematical Sciences, Daejeon 305-390, Korea*
[131]*Université de Lyon, F-69361 Lyon, France*
[132]*Hobart and William Smith Colleges, Geneva, New York 14456, USA*
[133]*Janusz Gil Institute of Astronomy, University of Zielona Góra, 65-265 Zielona Góra, Poland*
[134]*King's College London, University of London, London WC2R 2LS, United Kingdom*
[135]*Andrews University, Berrien Springs, Michigan 49104, USA*
[136]*Università di Siena, I-53100 Siena, Italy*
[137]*Trinity University, San Antonio, Texas 78212, USA*
[138]*University of Washington, Seattle, Washington 98195, USA*
[139]*Kenyon College, Gambier, Ohio 43022, USA*
[140]*Cornell University, Ithaca, New York 14853, USA*
[141]*Caltech JPL, Pasadena, California 91109, USA*

[a]Deceased, March 2016.
[b]Deceased, May 2015.
[c]Deceased, March 2015.